\documentclass[letterpaper, 10 pt, conference]{ieeeconf}  
\usepackage{xcolor}

\usepackage{cite}
\usepackage{amsmath,amssymb,amsfonts}
\usepackage{algorithm,algorithmic}
\usepackage{graphicx}
\usepackage{textcomp}
\usepackage{booktabs}
\usepackage{multirow}
\usepackage{multicol}
\usepackage{bm}
\usepackage{hyperref}
\usepackage{url}
\usepackage{subfig}
\usepackage[normalem]{ulem}

\IEEEoverridecommandlockouts                              

\title{\LARGE \bf
Learning and Optimization for Price-based Demand Response of Electric Vehicle Charging}

\author{Chengyang Gu$^{1}$, Yuxin Pan$^{2}$, Ruohong Liu$^{1}$ and Yize Chen$^{1}$
\thanks{$^{1}$ AI Thrust, Information Hub, The Hong Kong University of Science and Technology (Guangzhou). {\tt\small \{cgu893,rliu519\}@connect.hkust-gz.edu.cn , yizechen@ust.hk}}%
\thanks{$^{2}$ Division of Emerging Interdisciplinary Area under IPO, The Hong Kong University of Science and Technology. {\tt\small yuxin.pan@connect.ust.hk}}%
}%

\begin{document}

\maketitle
\thispagestyle{empty}
\pagestyle{empty}

\begin{abstract}
In the context of charging electric vehicles (EVs), the price-based demand response (PBDR) is becoming increasingly significant for charging load management. Such response usually encourages cost-sensitive customers to adjust their energy demand in response to changes in price for financial incentives. Thus, to model and optimize EV charging, it is important for charging station operator to model the PBDR patterns of EV customers by precisely predicting charging demands given price signals. Then the operator refers to these demands to optimize charging station power allocation policy. The standard pipeline involves offline fitting of a PBDR function based on historical EV charging records, followed by applying estimated EV demands in downstream charging station operation optimization. 
In this work, we propose a new decision-focused end-to-end framework for PBDR modeling that combines prediction errors and downstream optimization cost errors in the model learning stage. We evaluate the effectiveness of our method on a simulation of charging station operation with synthetic PBDR patterns of EV customers, and experimental results demonstrate that this framework can provide a more reliable prediction model for the ultimate optimization process, leading to more effective optimization solutions in terms of cost savings and charging station operation objectives with only a few training samples.

\end{abstract}

\section{INTRODUCTION}

Recent fast-growing penetration of electric vehicles (EVs) casts unforeseen challenges for charging station and power grid operations. Massive number of charging sessions inevitably result in overload due to the cluster effect on excessively concentrated EV charging stations and charging times~\cite{hadley2009potential, li2024diffcharge}. As both charging station and EV usually have limited capacities, excessive charging requests during peak times can cause considerable operational challenges. It is thus hard for charging stations to both satisfy consumers’ demands and maximize charging profits.





Thus, it is desirable to alleviate the overload of the EV charging stations with appropriate charging control. And when optimizing the charging control policy, one non-negligible factor is the price-based demand response (PBDR) of EV chargers. Cost-sensitive EV customer can actively adjust their total amount of charging energy based on the assigned charging prices~\cite{cedillo2022dynamic}. Therefore, in order to better operate simultaneous EV charging sessions, the price-based demand response (PBDR) strategy needs to be investigated through modeling the inherent relationship between energy demands and price~\cite{shao2011demand, bitar2016deadline, conejo2010real, yao2016real, xu2014operation}. Knowing such PBDR patterns of EV customers, the charging station operator can precisely estimate customers' charging requirements and hence develop reasonable station-level charging policy to minimize operation cost while satisfying charging demands.   

Previous research attempts utilized price-based demand response to encourage EV charging flexibility~\cite{develder2016quantifying}. 
\cite{bitar2016deadline} turned to a rule-based approach to design the deadline-differentiated electric power service. 
Yet such rule-based methods heavily rely on a set of pre-defined heuristics or pre-assumed conditions. 
An on-off EV demand response strategy was leveraged by formulating a binary optimization problem~\cite{yao2016real}, and finite-horizon Markov decision process was used to coordinate consumer owned energy storage~\cite{xu2014operation}. These methods however simplified the demand response model such that \emph{charging demands were assumed known exactly beforehand}. 
Moreover, both rule-based and optimization-based methods do not take into account the modeling procedure of price-based demand, and the charging demand is either estimated or assumed known exactly. An inaccurate forecast of price responsive charging load would further incur loss of robustness and reliability of station operation~\cite{donti2017task}. In this paper, we take a specific look into the relationship between charging price assigned to each EV owner and EV's charging demand, and investigate how to optimize charging strategy when such relationship is unrevealed to the charging station. We answer the following research question:
\newline \emph{How do we optimize EV charging scheduling over unknown demands incurred by variable charging price?}

\begin{figure}
    \centering
    \includegraphics[width=0.44\textwidth]{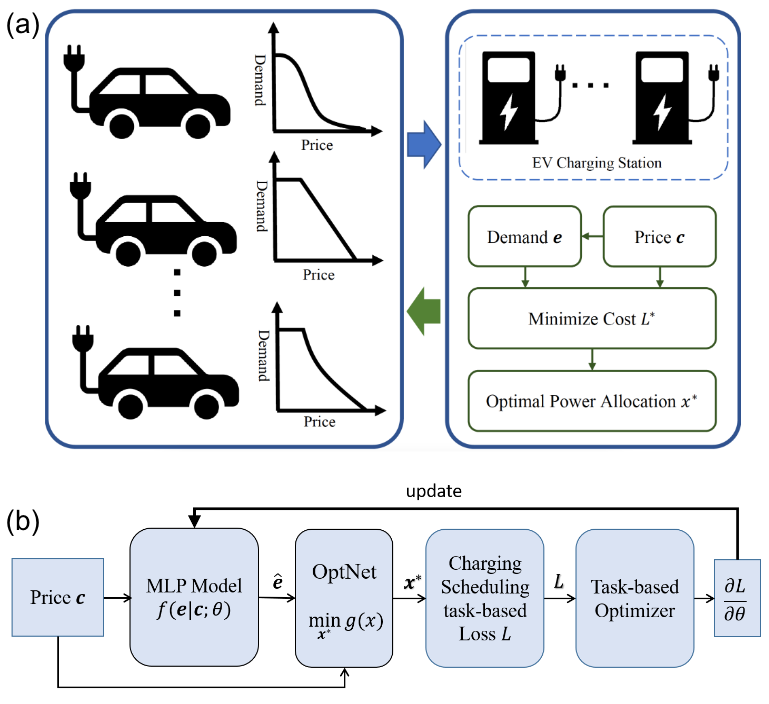}
    \caption{ (1a). Proposed EV demand estimation and charging station operation model. EV customers have various patterns of demand response to price. 
    \quad (1b). Proposed end-to-end charging demand forecasts and charging scheduling framework. 
    With optimization task-based loss $\mathcal{L}$, our model propagates loss derivative with respect to neural network model's parameter $\theta$ to help train the prediction model $f(\cdot)$. \vspace{-30pt}}
    \label{fig:1}
\end{figure}

We tackle both the demand forecasting and multi-step charging optimization problems in one framework. 
The proposed paradigm holds the promise of adaptively improving its performance over time when exposed to more data and feedback, making it well-suited for handling complex and dynamical problems. 
Standard pipelines involve individually fitting of a demand response function using the collected EV charging data, followed by the incorporation of predicted results into the ultimate optimization task~\cite{rockafellar1991scenarios, antonopoulos2021data}. Such workflow neglects the fact that the locally accurate predictions may not benefit the underlying optimization problem's overall objectives~\cite{donti2017task,amos2017optnet}. Moreover, the complex physical constraints make such predict-then-optimize framework produce infeasible solutions.  \cite{bian2022demand} considered demand response model in an end-to-end fashion, while the application was limited to identifying a subset of demand response model parameters rather than solving the optimization problem.

Our formulation makes it possible to backpropagate from the ultimate charging task-based objective to the learning of individual PBDR model. Moreover, it elicits the control actions for EV charging. The schematic of proposed method is shown in Fig. \ref{fig:1}.  Specifically, to find the price-responsive EV charging demands, we learn the complex price-based demand response (PBDR) as a function of charging price unrevealed to the system operator. We combine prediction errors and downstream optimization cost errors in the model learning stage. Experimentally, our method is evaluated on noisy demand response samples based on varying charging prices. The results demonstrate that our framework can lead to a considerable decrease in the operation costs by more than $20\%$ compared to the standard two-step (predict-then-optimize) method. This advantage is even more significant when the training dataset is small, meaning only limited customer responses are revealed to the charging stations.

\section{Problem Formulation}

In this section, we describe the formulation of charging station operation problem, in which we further develop the relationship between the electricity selling price assigned to EV customers and their corresponding total electricity demand as a price-based demand response (PBDR) model.
\subsection{Charging Station Operation Model}
We first formulate the charging station operation model. We suppose the goal of one EV charging station is to find the optimal policy to minimize its operation cost, meanwhile satisfying its customers' charging demands as much as possible. Each EV comes to the charging station at various timesteps, and reveals their requested energy based on the charging price (costs in per kWh) along with the departure time. The charging station can control the charging rate of each EV at each timestep. Denote $N$ as the number of active EV customers at the charging station, $1 \leq t \leq T$ as the station's opening time horizon. Then the charging station scheduling problem is as follows:

\begin{subequations}
\begin{align}
\label{equOptimize}
\nonumber \min _{\mathbf{x}} \;  g(\bm{x})= & \sum_{t}p(t)\sum_{i} x_{i}(t) - \sum_{i} c_i\sum_{t} x_{i}(t) +\\ & \beta\big(\sum_{i} (\sum_{t}x_{i}(t) -e_{i})^2\big) + \alpha \sum_{i, t} x_{i}^{2}(t)  \\
\text { s.t. } \quad & 0 \leq \sum_{i} x_{i}(t) \leq \overline{x}, \; \forall t\\
& 0 \leq  x_{i}(t) \leq \Bar{y}_{i}, \; \forall i, t\\
\label{equArrival}
& x_{i}(t) = 0, \; for \  t \leq t_{a,i} \ and \  t \geq t_{d,i}\\
& e_i = f_i(c_i);
\end{align}
\end{subequations}
where $\bm{x}=[x_{1}(1),x_{1}(2),...,x_{1}(T) ,x_{2}(1),..., ,x_{N}(T)]^{T}$ is the control variable; $x_i(t)$ represents the charging power assigned to EV customer $i$ at time $t$; $p(t)$ is the electricity purchase price from the grid; $c_i$ and $e_i$ are the electricity selling price and total electricity demand for EV customer $i$. $\overline{x}$ is maximum charging power at time $t$ for the station. $\Bar{y}_{i}$ is the maximum charging power for EV customer $i$. With arrival time $t_{a,i}$ and departure time $t_{d,i}$, and $\beta, \alpha$ is the coefficient for regularizing demand charging completion and non-smooth charging power.

With known $e_i$, this EV scheduling problem is a constrained Quadratic Programming (QP) problem which strong duality holds, and it can be solved to find optimal charging policy. However, it can be challenging to solve in practice with unknown $e_i$. Since each EV owner has distinct charging behaviors, while the charging demand can be varying with respect to price. From the charging station side, it cannot solve \eqref{equOptimize} without knowing such charging demand information. Customers adjust their electricity demands according to the selling price signal, and each customer may have its unique response to the price $e_i$. In order to model this price-demand map and use it to infer demand given price, we form the following demand price-response models. We note that in our setting, charging station operator informs customer $c_i$ before charging starts, and the optimization of $c_i$ along with charging scheduling is left for future work.

\begin{figure}[t]
    \centering
    \includegraphics[width=0.4\textwidth]{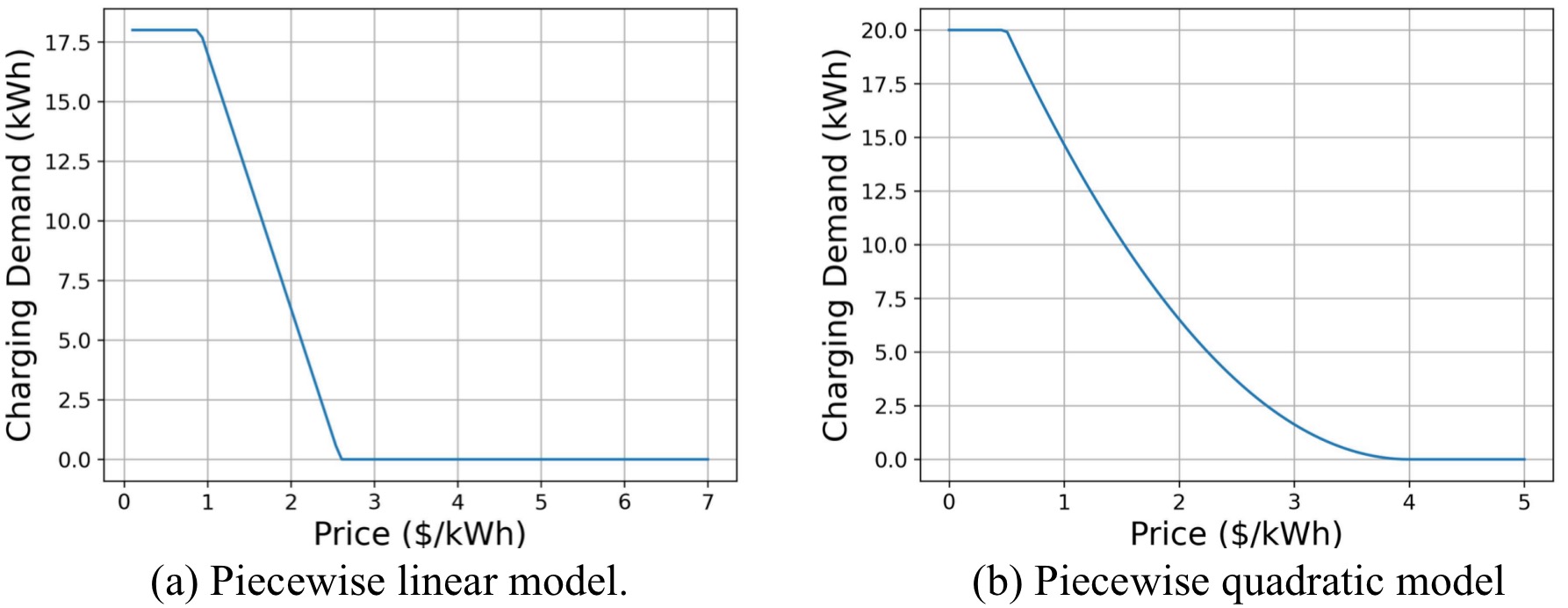}
    \caption{Examples of EV PBDR patterns. Individual EV's charging demand is a function of charging price.\vspace{-10pt}}
    \label{fig:patterns}
\end{figure}

\subsection{Price-based Demand Response Model}
In this work we include two examples on the explict modeling of price-based EV demand response which is not considered in previous research. We note that our solution techniques described in Section \ref{sec:method} are not limited to the particular function mapping from charging price to charging demand, and we are able to solve the charging station optimization problem under various unknown price-responsive demand patterns.

\textbf{Example 1: Utility-maximization EV demand}

Once the EV owner $i$ comes to the charging station and gets the price $c_i$ revealed, the EV aims at minimizing the charging costs while maximizing the charging utility, given the prevailing electricity price. The utility function $f(e_i)$ is usually assumed to be continuous, nondecreasing, and concave \cite{li2012optimization}. In this paper, the user utility function is modeled as $f(e_i) = -(e_{level,i}-e_{trip,i})^2$, where $e_{level,i} = e_{init,i}+e_i \eta_i$. $e_i, e_{level,i}, e_{trip,i}$, and $ e_{init,i}$ denote charging demand, final energy level, desired energy level after charging session, and initial energy level, respectively. Given that the users know their energy requirement for an upcoming trip, their utility function represents how well the charging need is met. Such utility maximization problem is formulated as
\begin{subequations}
\begin{align}
\label{equ:charging}
\min _{\mathbf{e_i}} \quad &   \gamma_1 c_i e_i + \gamma_2(e_{level,i}-e_{trip,i})^2 \\
\text { s.t. }  \quad & e_{level,i} = e_{init,i}+e_i \eta_i \label{equ:charging_limit}\\
& SOC_{min,i}\leq e_{level,i} \leq SOC_{max,i}\label{equ:charging_SOC}\\
& \frac{e_{i}}{(t_{d,i}-t_{a,i})} \leq p_{max,i} \label{equ:charging_average}
\end{align}
\end{subequations}

In the formulation above,  Constraint \eqref{equ:charging_limit} states that the energy level $e_{level,i}$ of the $i^{th}$ electric vehicle (EV) is equal to the sum of its initial energy level $e_{init,i}$ and the charging demand $e_i$; \eqref{equ:charging_SOC} requires that the energy level $e_{level,i}$ must fall within the battery state-of-charge capacity range; \eqref{equ:charging_average} specifies that the average charging power rate must not exceed the maximum power rate. $\gamma_1$ and $\gamma_2$ denote the weighting factors of the charging cost and utility term; $\eta_i \in (0,1)$ represents the charging efficiency. Generally, the charging demand $e_i$ obtained by solving \eqref{equ:charging} will have a piece-wise linear relationship with price signal $c_i$. Fig. (\ref{fig:patterns})(a) illustrates one example of such pattern.

\textbf{Example 2: Piecewise quadratic EV demand}

Some EV  owners are price-sensitive over certain price range, meaning that an increase in the charging cost results in a rapid decline in their demand for charging. To approximate the behaviour of this kind of customers, we adopt a general price-demand model and construct the piecewise function with quadratic components for their price-response model:
\begin{equation}
    e_i = \left\{
	\begin{aligned}
    & e_{i,base}, \text{ if $c_i \leq c_{i,s}$}\\
    & \max (e_{i,base} - \mu_i (c_i - c_{i,s})^2, 0), \text{ if $c_i > c_{i,s}$}
    \end{aligned}
    	\right
 .
\end{equation}
where $e_{i,base}$ denotes the original charging demand of the $i^{th}$ EV; $\mu_i$ represents the price sensitive parameter; when the current price $c_i$ is greater then its sensitive price level $c_{i,s}$, the EV user decides to decrease charging demand. Fig. (\ref{fig:patterns})(b) illustrates one example of demand price-response model based on the definition mentioned above.

\section{End-to-End Learning and Optimization}
\label{sec:method}
In practice, charging station is faced with unknown price-based demand response model for each customer. It is challenging to design charging control under such situation. In this section, we illustrate how it is possible to include the training of price-demand forecasting module as a fully differentiable part in the EV charging optimization model. 

\subsection{Reformulation of Charging Station Model}
To simplify the analysis, we first reformulate the optimization problem, and rewrite the charging station optimization problem 
\eqref{equOptimize} as a standard Quadratic Programming (QP):

\begin{small}
\begin{subequations}
\label{equ:QP}
\begin{align}
\label{QPObjective}
\nonumber \min _{\mathbf{x}}  g(\bm{x})= &\beta(B\bm{x}-\hat{\bm{e}})^{T}(B\bm{x}-\hat{\bm{e}})+(\bm{p}^{T}-\bm{c}^{T}B)\bm{x} + \alpha \bm{x}^{T}\bm{x}\\
\nonumber = &\frac{1}{2} \bm{x}^T(2\beta B^{T}B+2\alpha I)\bm{x} + (\bm{p}^{T}-\bm{c}^{T}B-2\beta \hat{\bm{e}}^{T}B) \bm{x}\\
 &\; +\beta \hat{\bm{e}}^{T}\hat{\bm{e}} \\ 
\text { s.t. } & \left[
 \begin{matrix}
   -A  \\
   A   \\
   -I   \\
   I
  \end{matrix}
  \right] \bm{x}  \leq \left[
 \begin{matrix}
   \bm{0}  \\
   \bm{r}   \\
   \bm{0}   \\
   \bm{y}
  \end{matrix}
  \right]  \\ 
& F\bm{x} = \bm{0}.
\end{align}
\end{subequations}
\end{small}

\begin{itemize}
    \item $\bm{p}=[p(1), p(2), ..., p(T), ... ... , p(1), p(2), ..., p(T)]^{T} \in R^{NT \times 1}.$ lists the electricity price for charging $x_{i}(t)$.
    \item $\bm{c}=[c_1, c_2, .., c_N]^{T}$ lists the selling price assigned to each EV customer $i$.
    \item $\hat{\bm{e}}=[\hat{e}_1, \hat{e}_2, .., \hat{e}_N]^{T}$ lists the estimated demand of each EV customer $i$ by a Multi-layer Perceptron (MLP) model $\bm{\hat{e}} = f(\bm{c}; \theta)$. 
    \item $\bm{r} =\Bar{x}\bm{1}_{(T \times 1)} \in R^{T \times 1}$ represents the charging capacity (\textit{i.e.} station level maximum charging power).
    \item $\bm{y} = [\Bar{y}_1\bm{1}^{T}_{(T \times 1)}, \Bar{y}_2\bm{1}^{T}_{(T \times 1)}, ..., \Bar{y}_N\bm{1}^{T}_{(T \times 1)}]^{T} \in R^{NT \times 1}$ represents the customer level maximum charging power.
    \item $A=[I_{(T \times T)}, ..., I_{(T \times T)}] \in R^{T \times NT}$ helps transform term $\sum_{i} x_{i}(t)$ into matrix form. $\sum_{i} x_{i}(t)=A\bm{x}$.
     \item $B=\left[\begin{matrix}
   \bm{1}^{T}_{T \times 1} & 0  & ... & 0  \\
   \ 0 & \bm{1}^{T}_{T \times 1} &... & 0  \\
   \ ... & ... &... & ...  \\
   \ 0 & 0 &... & \bm{1}^{T}_{T \times 1}
  \end{matrix}\right] \in R^{N \times NT}$ helps transform term $\sum_{i}
  (\sum_{t}x_{i}(t) -\hat{e}_{i})^2$ into matrix form. $\sum_{i} (\sum_{t}x_{i}(t) -\hat{e}_{i})^2 = B\bm{x} - \bm{\hat{e}}$.
    \item $F \in R^{H \times NT}$ transforms the arrival and departure constraints \eqref{equArrival} into matrix form $F\bm{x} = \bm{0}$. $H$ is the total number of elements $x_{i}(t)$ which have constraint $x_i(t)=0$.  Each line $F_{h}$ of $F$   corresponds to one particular equality constraint $x_{i_h}(t_h)=0$; in this line only element $F_{h, i_{h} \times T + t_{h}} = 1$.
   
\end{itemize}

Previous works usually suppose that if we have an accurate estimation $\hat{\bm{e}}$ of actual demand $\bm{e}$, the system cost \eqref{QPObjective} will also achieve its optimal~\cite{amini2016arima}. Hence, two-step approach separates prediction and optimization: first to train the MLP $f(\bm{c}; \theta)$ by minimizing root mean squared error (RMSE) between customer demand prediction and actual charging demand; then to use the trained MLP to predict customer demands $\hat{\bm{e}}$, and based on the predicted demands they solve problem \eqref{equ:QP} to obtain minimum operation cost.

Our goal is to learn the optimal parameter $\theta$ of an MLP prediction model $\bm{\hat{e}}=f(\bm{c}; \theta)$, which approximates customer's demand response to price signal. Based on it we can minimize \textbf{actual} charging station operation cost $\mathcal{L}(\theta)$: 
\begin{equation}
    \label{LossOptimizer}
    \begin{aligned}
         \min_{\theta} \mathcal{L}(\theta)= \beta(B\bm{x}^{*}(\theta)-\bm{e})^{T}(B\bm{x}^{*}(\theta)-\bm{e})& \\
         +(\bm{p}^{T}-\bm{c}^{T}B)\bm{x}^{*}(\theta)+ \alpha \bm{x}^{*}(\theta)^{T}\bm{x}^{*}(\theta);
    \end{aligned}
\end{equation}
in which $\bm{x^{*}}(\theta)$ represents the optimal solution of the charging station operation optimization problem \eqref{equ:QP}. $\bm{e}$ denotes the actual demand of each EV customer.

In contrast to the existing paradigm, our learning model is decision-focused, which means in the learning process we focus on the downstream optimization performance rather than solely on the prediction accuracy. The overall algorithm is an end-to-end framework to find the optimal decisions with price signals as input. Fig. (\ref{fig:1}) illustrates the architecture of our framework. More specifically, we directly use the objective function \eqref{LossOptimizer}  of the charging station operation model as the training loss for the demand forecasting model MLP $f(\cdot)$, rather than the RMSE of demand prediction, and back propagate its gradient with respect to MLP model parameters $\frac{\partial \mathcal{L}}{\partial \theta}$, to adapt the model: $    \frac{\partial \mathcal{L}}{\partial \theta} = \frac{\partial  \mathcal{L}}{\partial \bm{x^{*}}} \frac{\partial \bm{x^{*}}}{\partial \theta}.$ From \eqref{LossOptimizer} we can easily obtain $\frac{\partial  \mathcal{L}}{\partial \bm{x^{*}}}$. Hence, we focus on computing the partial derivative $\frac{\partial \bm{x^{*}}}{\partial \theta}$ hereafter. 

\subsection{Differentiating the Optimal Solution}

Now we analyze the optimal solution $\bm{x}^{*}$ of this standard form QP problem \eqref{equ:QP}. Since strong duality holds, $\bm{x}^{*}$ should satisfy the Karush-Kuhn-Tucker (KKT) Conditions:
\begin{small}
    \begin{subequations}
    \label{KKT}
    \begin{align}
       \nonumber (2\beta B^{T}B+2\alpha I) \bm{x}^{*} + &(\bm{p}^{T}-\bm{c}^{T}B-2\beta \hat{\bm{e}}^{T}B)^{T} + \left[
 \begin{matrix}
   -A  \\
   A   \\
   -I   \\
   I
  \end{matrix}
  \right]^{T} \bm{\lambda}^{*} \\
  + F^{T}\bm{\nu}^{*} = 0  \\
    &F\bm{x}^{*} = 0\\
    diag(\bm{\lambda}^{*}) &(\left[
 \begin{matrix}
   -A  \\
   A   \\
   -I   \\
   I
  \end{matrix}
  \right]  \bm{x}^{*} - \left[
 \begin{matrix}
   \bm{0}  \\
   \bm{r}   \\
   \bm{0}   \\
   \bm{y}
  \end{matrix}
  \right]  ) = 0
    \end{align}
\end{subequations}

\end{small}
\noindent where $\bm{\lambda}^{*}$ and $\bm{\nu}^{*} $ are optimal dual variables. We compute the target partial derivative $\frac{\partial \bm{x^{*}}}{\partial \theta}$ by differentiating Equation \eqref{KKT}. The matrix form of the KKT differentials is:
\begin{scriptsize}
    \begin{equation}
    \label{KKTDifferentials}
    \underbrace{\left[\begin{matrix}
        2\beta B^{T}B+2\alpha I & [-A^{T} A^{T} -I \quad I] & F^{T}\\
        diag(\bm{\lambda}^{*})\left[
 \begin{matrix}
   -A  \\
   A   \\
   -I   \\
   I
  \end{matrix}
  \right] &  diag(\left[
 \begin{matrix}
   -A  \\
   A   \\
   -I   \\
   I
  \end{matrix}
  \right]  \bm{x}^{*} - \left[
 \begin{matrix}
   \bm{0}  \\
   \bm{r}   \\
   \bm{0}   \\
   \bm{y}
  \end{matrix} 
  \right]  ) & 0 \\
 F & 0 & 0
    \end{matrix}\right] }_{G}
    \left[\begin{matrix}
    d\bm{x}^{*} \\
    d\bm{\lambda}^{*} \\
    d\bm{\nu}^{*} \\
    \end{matrix}\right] =  \left[\begin{matrix}
    2\beta B^{T} d\bm{\hat{e}} \\
    \bm{0}\\
    \bm{0} \\
    \end{matrix}\right]
\end{equation}
\end{scriptsize}

We divide both sides of Equation \ref{KKTDifferentials} with differential $d \theta$. And then we have a set of linear equations to solve for target derivative $\frac{\partial \bm{x^{*}}}{\partial \theta}$:
\begin{small}
    \begin{equation}
    \label{KKTDifferentials2}
    G
    \left[\begin{matrix}
    \frac{d\bm{x}^{*}}{d\theta} \\
    \frac{d\bm{\lambda}^{*}}{d\theta} \\
    \frac{d\bm{\nu}^{*}}{d\theta} \\
    \end{matrix}\right] =  \left[\begin{matrix}
    2\beta B^{T} \frac{d\bm{\hat{e}}}{d\theta} \\
    \bm{0} \\
    \bm{0} \\
    \end{matrix}\right]
\end{equation}
\end{small}

Matrix G is obviously an upper triangular matrix here. Hence, it will always be invertible. Linear Equation \ref{KKTDifferentials2} will achieve the unique solutions of gradients  $\frac{d\bm{x}^{*}}{d\theta}$,$\frac{d\bm{\lambda}^{*}}{d\theta}$, $\frac{d\bm{\nu}^{*}}{d\theta}$. This gives the guarantee to make the gradients and parameter updates to be well-defined. We refer to \cite{barratt2018differentiability} for a more detailed discussion on the model differentiability.

\subsection{Model Learning}
In practice, we can solve QP problem \eqref{equ:QP} in batches efficiently in deep learning environment like Pytorch. We adapt the neural network architecture defined in OptNet\cite{amos2017optnet}. Our approach is inspired by such end-to-end differentiable optimization framework \cite{amos2017optnet, donti2017task}. 

In our model, we output optimal solution $\bm{x}^{*}$ and compute the charging scheduling task-based loss gradient $\frac{\partial \mathcal{L}}{\partial \theta}$ by solving Equation \eqref{KKTDifferentials2}. Then we backpropagate the gradient to update parameter $\theta$ in MLP model $f(\bm{\hat{e}}|\bm{c}; \theta)$ until the loss converges to an optimal value. Once the model is trained, we use it to generate estimated demand $\hat{\bm{e}}$. Then we forward $\hat{\bm{e}}$ to the differentiable optimization module, and it will solve \eqref{equ:QP} and output optimal solution $\bm{x}^*$. Given $\bm{x}^*$ we use \eqref{LossOptimizer} to compute the actual optimal operation cost $\mathcal{L}$.
The whole end-to-end model learning process is illustrated in Algorithm. \ref{Algorithm}.

\begin{algorithm}[h]
\caption{Charging Scheduling task based loss optimization}
\label{Algorithm}
\begin{algorithmic}[1]
\STATE \textbf{Input:} price signal $\bm{c}$
\STATE \textbf{Initialize} parameter $\theta$ for MLP model $f(\bm{\hat{e}}|\bm{c};\theta)$.
\FOR{Iteration $j$ ($j = 1, 2,..., J_{max}$) } 
\STATE Estimate $\hat{\bm{e}}=f(\bm{c}|\theta)$.
\STATE Solve Equation \eqref{equ:QP} for optimal solution $\bm{x}^{*}$ with estimated $\hat{\bm{e}}$.
\STATE With $\bm{x}^{*}$ compute derivative $\frac{\partial \mathcal{L}(\bm{x}^{*})}{\partial \theta}$ by solving Equation \eqref{KKTDifferentials2}.
\STATE Update $\theta_{j+1} \leftarrow \theta_{j} - \eta \frac{\partial \mathcal{L}}{\partial \bm{x}^{*}} \frac{\partial \bm{x}^{*}}{\partial \theta} $.
\ENDFOR
\end{algorithmic}
\end{algorithm}

\vspace{-10pt}\section{Numerical Simulations}
In the simulation, we consider running a real-world charging station with electricity price data in Anaheim City, California~\cite{Purchase}, which has limited charging capacity and serves multiple EV customers for each day. 
We evaluate the performance with the overall actual optimal operation cost $\mathcal{L}(\theta)$, and 3 sub evaluation metrics: (1) \textit{operation profit} $(\bm{p}^{T}-\bm{c}^{T}B)\bm{x}^{*}$, representing the profits earned by charging station. (2) \textit{charging demand completion penalty} $ \beta(B\bm{x}^{*}-\hat{\bm{e}})^{T}(B\bm{x}^{*}-\hat{\bm{e}})$, reflecting the demand satisfaction rate of EV customers. Lower penalty value represents higher demand satisfaction rate. (3) \textit{smooth penalty} $\alpha (\bm{x}^{*})^{T}\bm{x}^{*}$, reflecting the smoothness of charging curves under charging policy $\bm{x}^{*}$. 
We compare our framework with ground-truth results ($i.e.$  solve \eqref{equ:QP} for optimal solution $\bm{x}^*$ under no demand estimation errors) and the two-step approach (as compared baseline). All models are built and trained with NVIDIA RTX 2080 Ti and Intel(R) Xeon(R) Silver 4210 CPU.


\begin{figure}[t]
  \centering
  \subfloat[]{%
    \includegraphics[width=0.23\textwidth]{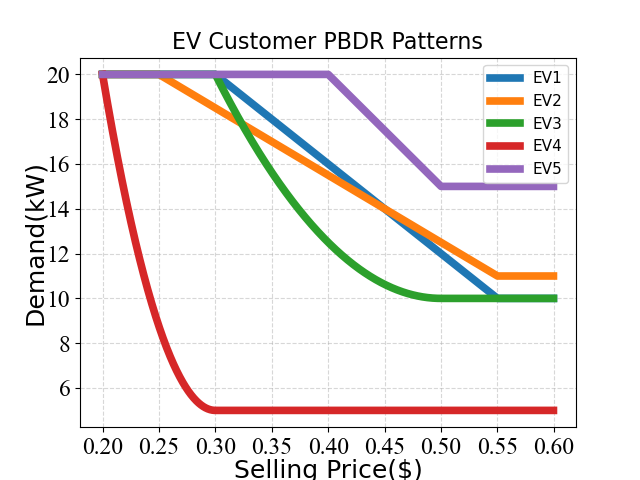}%
    \label{fig:subfig1}%
  }
  \subfloat[]{%
    \includegraphics[width=0.23\textwidth]{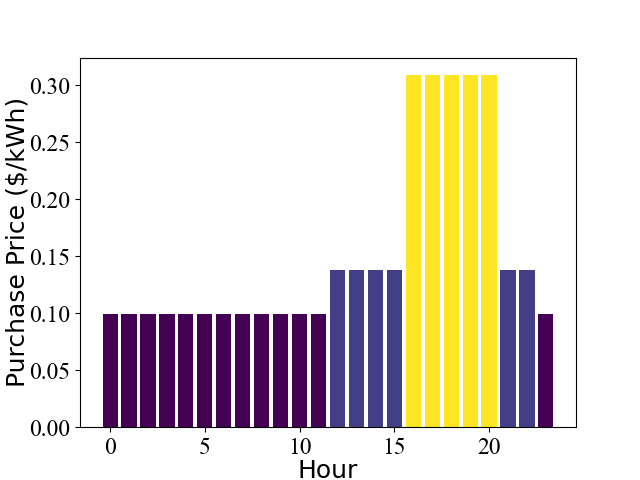}%
    \label{fig:subfig2}%
  }
  \caption{(\ref{fig:subfig1}) shows the synthetic price-based demand response (PBDR) patterns used in the simulation; 
  (\ref{fig:subfig2}) shows the real-world hourly commercial electricity purchase price \cite{Purchase} with colors indicating different price-time ranges. We set this peak-valley purchase price as $\bm{p}$ in our simulation.\vspace{-10pt}}
  \label{fig:mainfig}
\end{figure}

In the charging station model, we assume that the operational horizon of station is $T=24$ hours since EV charging follows daily commute patterns, and the number of EV customers $N=5$. Each EV customer can charge at a particular station at different starting times, so each EV customer has a unique charging session. 
We assign $3$ Level-2 charging ports with $6.6$ kW maximum charging power and $2$ Level-2 charging ports with $3.6$ kW maximum charging power for all $5$ EV customers. The penalty coefficient for charging demand completion is $\beta=5$, and the penalty coefficient for smooth is $\alpha=0.001$. 
We construct 3 Piece-wise Linear and 2 Piece-wise Quadratic PBDR patterns as the synthetic PBDR patterns of $5$ EV customers in the simulation based on  Section \textit{II.B}'s models (see Fig. (\ref{fig:subfig1})). We set all charging demands of EVs varying in range [5, 20] (kW) based on general charging cases in Caltech ACN Dataset\cite{lee2019acndata}. We also simulate customers with varying price sensitivity parameters and price sensitivity ranges. Based on the synthetic PBDR patterns, we generate a training dataset of 1,000 samples and a testing dataset of 100 samples. For the  selling price $\bm{c}$, we test two settings, where we 
assign random and different prices to 5 different EV customers \cite{DiffPrice}. Our method works well under both settings, and the price range [0.2, 0.6] (\$/kWh) is based on latest market price in U.S. 2023 \cite{Sell}. Then we use synthetic PBDR patterns to compute corresponding actual demand response $\bm{e}$. In this dataset we treat $\bm{c}$ as feature, $\bm{e}$ as label. We use real-world electricity purchase price shown in Fig. (\ref{fig:subfig2}) for the optimization.

\begin{figure}[]
  \centering
  \subfloat[]{%
    \includegraphics[width=0.23\textwidth]{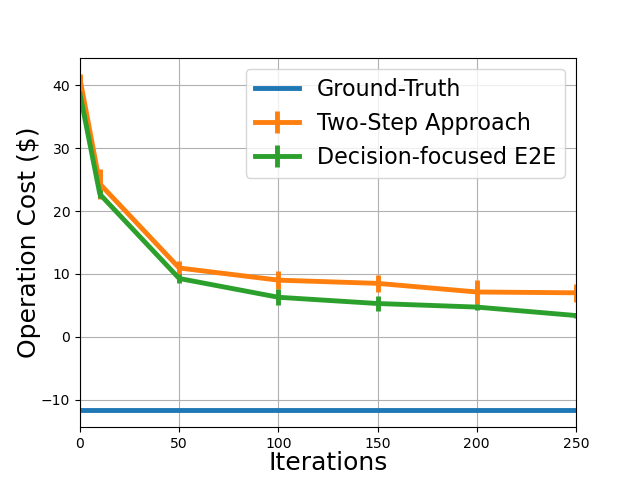}%
    \label{fig:subresult1}%
  }
  \subfloat[]{%
    \includegraphics[width=0.23\textwidth]{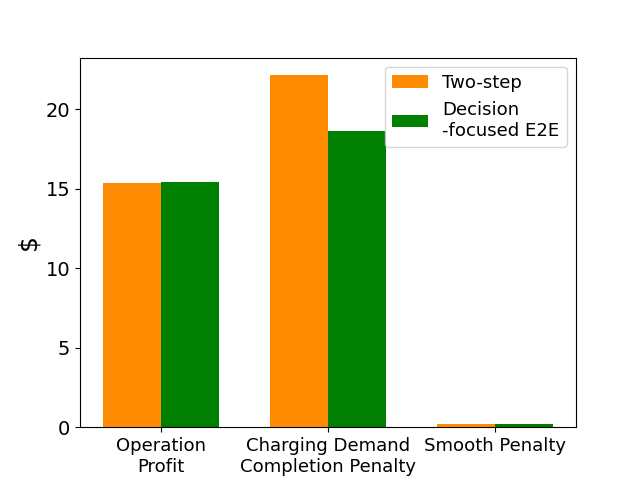}%
    \label{fig:subresult2}%
  }
  \caption{(a) operation costs during training, results averaged over 5 runs; (b) comparisons of sub-evaluation metrics.\vspace{-15pt}}
  \label{fig:Result}
\end{figure}

In the numerical simulation, we examine how different methods minimize the actual operation cost $\mathcal{L}$ of the established charging station operation model with settings in Section \textit{IV.A}. To benchmark our method, we develop both two-step approach and our decision-focused end-to-end framework (decision-focused E2E). Same as the network architecture applied in \cite{donti2017task}, we employ a commonly-used 2-hidden-layer network with a residual connection between inputs and outputs as the MLP model in both two-step approach and our decision-focused end-to-end framework. We respectively use 200, 400, 600, 800 samples in training dataset generated in Section \textit{IV.B} to train the MLP model.  For both methods we train the models for 250 iterations. We evaluate the performance of two methods with the average operation cost, profit, penalties of demand completion and smooth on the 100-sample test dataset in Section \textit{IV.B}. All the results are benchmarked against ground-truth setting with known individual EV's charging price and demands.

\begin{table}[H]
    \centering
    \caption{Evaluation on operation cost $\mathcal{L}$ (\$) with different number of training samples. Ground-Truth value is -11.69.\vspace{-8pt}}
    \begin{tabular}{c|c|c|c}
           \hline
       \textbf{Samples} & \textbf{Decision-focused E2E} & \textbf{Two-step} &  \textbf{Improvement(\$)}\\ 
         \hline
         200 & 3.39$\pm$0.45& 7.01$\pm$1.44 & 3.62\\

          400 & -4.90$\pm$0.45 & -1.91$\pm$0.25 &  2.99\\

         600 & -7.26$\pm$0.24 & -4.45$\pm$0.59 & 2.81\\

         800 & -8.07$\pm$1.02 & -5.56$\pm$0.31 & 2.51\\
         \hline
    \end{tabular}
    \label{Performance}
\end{table}

Fig. (\ref{fig:subresult1}) illustrates the convergence of station's operation costs during 250 training iterations. The converged costs for both methods still have a gap compared to ground truth, as limited samples result in imperfect demand estimations. But as we expected, our decision-focused end-to-end framework outperforms the two-step approach, for it always converges to a better actual operation cost by explicitly training the forecasting MLP with the differentiable loss. 

\begin{table}[H]
\centering
\caption{Evaluation on sub-metric \textit{Operation Profit}, \textit{Charing Demand Completion Penalty}, and \textit{Smooth Penalty} with varying training samples. (5 runs average)}
\setlength{\tabcolsep}{3pt}
\begin{tabular}{c|cc|cc|cc}
\hline
& \multicolumn{2}{c|}{Operation Profits}&\multicolumn{2}{|c|}{Completion Penalty}&\multicolumn{2}{|c}{Smooth Penalty}\\
\textbf{Samples} & \textbf{E2E} & \textbf{Two-step} & \textbf{E2E} & \textbf{Two-step} &\textbf{E2E} & \textbf{Two-step}  \\ \hline
200              & 15.41                                & 15.33    &18.62                                & 22.15&0.1851                               & 0.1861                        \\
400              & 15.42                                & 15.36       &10.33                                & 13.26&    0.1858                               & 0.1871                  \\
600              & 15.39                                & 15.34       &7.95             & 10.70&0.1850            & 0.1865                        \\ 
800              & 15.45                                & 15.40        & 7.19            &9.65 &0.1865           & 0.1870                   \\ \hline
\end{tabular}
\label{tab:profit}
\end{table}

Table \ref{Performance} records the results of operation cost comparison under different numbers of training samples (All results are average of 5 runs). Our decision-focused end-to-end framework has more reliable performance on minimizing overall operation cost $\mathcal{L}$. The improvement upon twp-step approach is much more apparent with small training dataset, indicating our framework performs better under the case of limited history. This demonstrates the value of incorporating the training of demand forecasting models into the overall optimization problem. In Table \ref{tab:profit}, we show the corresponding results of 3 sub evaluation metrics: \textit{operation profit}, \textit{charging demand completion penalty}, \textit{smooth penalty}. We can see that \textit{charging demand completion penalty} contributes most to our framework's leading over the two-step approach, while improvements on \textit{operation profit} and \textit{smooth penalty} are relatively mild, as illustrated in Fig. (\ref{fig:subresult2}). The major reason is that our framework optimizes the operation cost $\mathcal{L}$ by improving MLP estimations of charging demands $\hat{\bm{e}}$, and \textit{charging demand completion penalty} is the only sub-evaluation metric that contains a second-order term of $\beta \hat{\bm{e}}^{T}\hat{\bm{e}}$; meanwhile in the simulation we assign a charging demand completion penalty coefficient $\beta=5$. This makes \textit{charging demand completion penalty} the most influenced metric by our proposed framework. In our formulation, we can also conveniently tune the penalty coefficients $\alpha, \beta$ in optimization objectives in Equation \eqref{QPObjective} so that the demand prediction model can emphasize on particular task (e.g., maximizing operation profits, smoothing charging curves). Proposed method can strike a balance between shaping the charging curves and optimizing the charging needs.

\begin{figure}[t]
    \centering
    \includegraphics[width=0.3\textwidth]{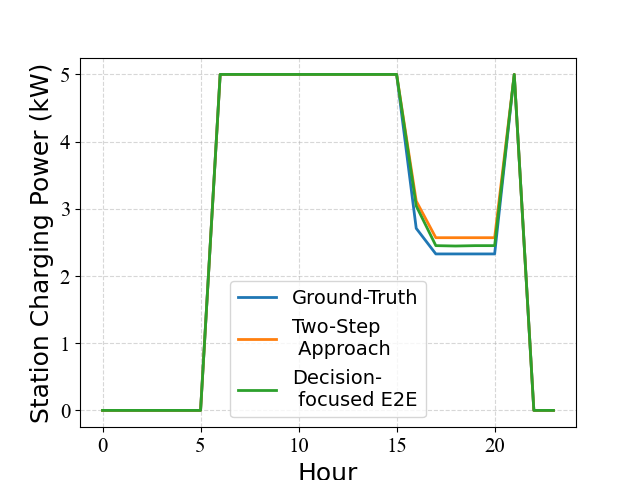}
    \caption{Comparison of two-step approach and our decision-focused end-to-end framework on station-level charging power (one example from test dataset). \vspace{-15pt}}
    \label{fig:stationpower}
\end{figure}

Fig. \ref{fig:subresult1} shows operation cost decreasing process of two-step approach and our decision-focused end-to-end framework; while Fig. \ref{fig:subresult2} shows our decision-focused end-to-end framework outperforms two-step approach much on \textit{charging demand completion penalty}.  In Fig. \ref{fig:stationpower}, we further look into the overall charging power delivered by the station. Compared to two-step approach, we can validate that our framework achieves lower total charging power, which is closer to ground-truth value. The major improvements occur in the peak-price period when both profit and completion rate metrics are crucial, and decision-makers need to find a balanced solution away from the boundaries to minimize the total cost. This result justifies that by taking the decision objective into forecasting modules, our framework can achieve higher session completion rate while consuming less total power with imperfect demand estimation, especially for periods that require a delicate balanced solution.

\vspace{-10pt}\section{Discussion and Conclusion}
In this paper, we propose an end-to-end learning and optimization framework for price-responsive EV charging session scheduling. In particular, we model the EV charging demands as unknown values in response to varying pricing signals for EV users.Simulation results validate that our proposed framework demonstrates superior performance compared to the existing two-step approach (prediction-based learning + optimizing on standalone prediction). 
Our advantage is particularly pronounced in cases with small training datasets, indicating the effectiveness of our framework in situations with limited historical charging records. In the future work, we will explore more efficient computation for KKT differentiation to enable the framework to tackle realistic charging problems with huge EV customers, and consider real-time implementation with stochastic EV arrivals. 
We will also try to enhance the generalization and robustness of EV PBDR modeling by training with more realistic, noisy datasets. 
Furthermore, we will explore the co-optimization of time-varying pricing assignments and charging schedules for individual customers. We are also interested in expanding the proposed end-to-end learning and optimization framework to encompass a broader range of demand response programs.


\bibliographystyle{IEEEtran}
\bibliography{bibfile}

\begin{thebibliography}{10}
\providecommand{\url}[1]{#1}
\csname url@rmstyle\endcsname
\providecommand{\newblock}{\relax}
\providecommand{\bibinfo}[2]{#2}
\providecommand\BIBentrySTDinterwordspacing{\spaceskip=0pt\relax}
\providecommand\BIBentryALTinterwordstretchfactor{4}
\providecommand\BIBentryALTinterwordspacing{\spaceskip=\fontdimen2\font plus
\BIBentryALTinterwordstretchfactor\fontdimen3\font minus
  \fontdimen4\font\relax}
\providecommand\BIBforeignlanguage[2]{{%
\expandafter\ifx\csname l@#1\endcsname\relax
\typeout{** WARNING: IEEEtran.bst: No hyphenation pattern has been}%
\typeout{** loaded for the language `#1'. Using the pattern for}%
\typeout{** the default language instead.}%
\else
\language=\csname l@#1\endcsname
\fi
#2}}

\bibitem{hadley2009potential}
S.~W. Hadley and A.~A. Tsvetkova, ``Potential impacts of plug-in hybrid
  electric vehicles on regional power generation,'' \emph{The Electricity
  Journal}, vol.~22, no.~10, pp. 56--68, 2009.

\bibitem{li2024diffcharge}
S.~Li, H.~Xiong, and Y.~Chen, ``Diffcharge: Generating ev charging scenarios
  via a denoising diffusion model,'' \emph{IEEE Transactions on Smart Grid},
  2024.

\bibitem{cedillo2022dynamic}
M.~H. Cedillo, H.~Sun, J.~Jiang, and Y.~Cao, ``Dynamic pricing and control for
  ev charging stations with solar generation,'' \emph{Applied Energy}, vol.
  326, p. 119920, 2022.

\bibitem{shao2011demand}
S.~Shao, M.~Pipattanasomporn, and S.~Rahman, ``Demand response as a load
  shaping tool in an intelligent grid with electric vehicles,'' \emph{IEEE
  Transactions on Smart Grid}, vol.~2, no.~4, pp. 624--631, 2011.

\bibitem{bitar2016deadline}
E.~Bitar and Y.~Xu, ``Deadline differentiated pricing of deferrable electric
  loads,'' \emph{IEEE Transactions on Smart Grid}, vol.~8, no.~1, pp. 13--25,
  2016.

\bibitem{conejo2010real}
A.~J. Conejo, J.~M. Morales, and L.~Baringo, ``Real-time demand response
  model,'' \emph{IEEE Transactions on Smart Grid}, vol.~1, no.~3, pp. 236--242,
  2010.

\bibitem{yao2016real}
L.~Yao, W.~H. Lim, and T.~S. Tsai, ``A real-time charging scheme for demand
  response in electric vehicle parking station,'' \emph{IEEE Transactions on
  Smart Grid}, vol.~8, no.~1, pp. 52--62, 2016.

\bibitem{xu2014operation}
Y.~Xu and L.~Tong, ``On the operation and value of storage in consumer demand
  response,'' in \emph{53rd IEEE Conference on Decision and Control}.\hskip 1em
  plus 0.5em minus 0.4em\relax IEEE, 2014, pp. 205--210.

\bibitem{develder2016quantifying}
C.~Develder, N.~Sadeghianpourhamami, M.~Strobbe, and N.~Refa, ``Quantifying
  flexibility in ev charging as dr potential: Analysis of two real-world data
  sets,'' in \emph{2016 IEEE International Conference on Smart Grid
  Communications}.\hskip 1em plus 0.5em minus 0.4em\relax IEEE, 2016, pp.
  600--605.

\bibitem{donti2017task}
P.~Donti, B.~Amos, and J.~Z. Kolter, ``Task-based end-to-end model learning in
  stochastic optimization,'' \emph{Advances in neural information processing
  systems}, vol.~30, 2017.

\bibitem{rockafellar1991scenarios}
R.~T. Rockafellar and R.~J.-B. Wets, ``Scenarios and policy aggregation in
  optimization under uncertainty,'' \emph{Mathematics of operations research},
  vol.~16, no.~1, pp. 119--147, 1991.

\bibitem{antonopoulos2021data}
I.~Antonopoulos, V.~Robu, B.~Couraud, and D.~Flynn, ``Data-driven modelling of
  energy demand response behaviour based on a large-scale residential trial,''
  \emph{Energy and AI}, vol.~4, p. 100071, 2021.

\bibitem{amos2017optnet}
B.~Amos and J.~Z. Kolter, ``{O}pt{N}et: Differentiable optimization as a layer
  in neural networks,'' in \emph{Proceedings of the 34th International
  Conference on Machine Learning}, ser. Proceedings of Machine Learning
  Research, vol.~70.\hskip 1em plus 0.5em minus 0.4em\relax PMLR, 2017, pp.
  136--145.

\bibitem{bian2022demand}
Y.~Bian, N.~Zheng, Y.~Zheng, B.~Xu, and Y.~Shi, ``Demand response model
  identification and behavior forecast with optnet: A gradient-based
  approach,'' in \emph{Proceedings of the Thirteenth ACM International
  Conference on Future Energy Systems}, 2022, pp. 418--429.

\bibitem{li2012optimization}
N.~Li, L.~Gan, L.~Chen, and S.~H. Low, ``An optimization-based demand response
  in radial distribution networks,'' in \emph{2012 IEEE Globecom
  Workshops}.\hskip 1em plus 0.5em minus 0.4em\relax IEEE, 2012, pp.
  1474--1479.

\bibitem{amini2016arima}
M.~H. Amini, A.~Kargarian, and O.~Karabasoglu, ``Arima-based decoupled time
  series forecasting of electric vehicle charging demand for stochastic power
  system operation,'' \emph{Electric Power Systems Research}, vol. 140, pp.
  378--390, 2016.

\bibitem{barratt2018differentiability}
S.~Barratt, ``On the differentiability of the solution to convex optimization
  problems,'' \emph{arXiv preprint arXiv:1804.05098}, 2018.

\bibitem{Purchase}
``Commercial rates. city of anaheim.''
  \url{https://www.anaheim.net/6440/Commercial-Rates}, accessed: 2023-09-11.

\bibitem{lee2019acndata}
Z.~Lee, T.~Li, and S.~H. Low, ``Acn-data: Analysis and applications of an open
  ev charging dataset,'' in \emph{Proceedings of the Tenth International
  Conference on Future Energy Systems (e-Energy '19)}, June 2019.

\bibitem{DiffPrice}
``The shocking truth: How the price of a charging station varies based on the
  type of vehicle being charged.'' \url{https://energy5.com/}, accessed:
  2023-09-27.

\bibitem{Sell}
``How much does it cost to charge an electric car?''
  \url{https://www.enelxway.com/us/en/resources/blog}, accessed: 2023-09-11.

\end{thebibliography}

\end{document}